\documentclass[conference,10pt]{IEEEtran}
\IEEEoverridecommandlockouts

\usepackage{amsmath}
\usepackage{eqparbox}
\usepackage[usenames, dvipsnames]{color}
\usepackage{steinmetz}
\usepackage{mathtools,amssymb,bm,mathabx,nccmath}
\usepackage{amstext}
\usepackage{amssymb,bm}
\usepackage{graphicx}
\usepackage{color}
\usepackage{booktabs}
\usepackage{longtable}
\usepackage{multicol}
\usepackage{algorithmicx}
\usepackage[ruled]{algorithm}
\usepackage{algpseudocode}
\usepackage{algpascal}
\usepackage{algc}
\usepackage{amsfonts}
\usepackage{dsfont}
\usepackage{array}
\usepackage[most]{tcolorbox}
\usepackage{cases}
\usepackage{pgfplots}
\pgfplotsset{compat=newest}
\usetikzlibrary{spy}
\usepackage{accents}
\usepackage{etoolbox}
\usepackage{caption}
\DeclareCaptionLabelSeparator{periodspace}{.~}
\usepackage{float}
\usepackage{subcaption}
\usepackage{xr-hyper}
\usepackage{hyperref}
\usetikzlibrary{3d}
\usepackage{amsthm}

\usetikzlibrary{positioning}
\usepackage{tikz-3dplot}
\usetikzlibrary{matrix}
\usepackage{float}
\usepackage{cite}
\usepackage{epstopdf}

\theoremstyle{remark}

\theoremstyle{plain}
\newtheorem{thm}{\textbf{Theorem}}
\newtheorem{lem}{\textbf{Lemma}}

\theoremstyle{definition}

\ifCLASSOPTIONcaptionsoff
\usepackage[nomarkers]{endfloat}
\let\MYoriglatexcaption\caption
\renewcommand{\caption}[2][\relax]{\MYoriglatexcaption[#2]{#2}}
\fi
\newtheorem{rem}{\textbf{Remark}}

\newcommand*{\rom}[1]{\expandafter\@slowromancap\romannumeral #1@}

\newcommand{\RN}[1]{%
\textup{\uppercase\expandafter{\romannumeral#1}}%
}

\usepackage{standalone}
\graphicspath{{./Figures/}}
\usepackage{times}
\usepackage[T1]{fontenc}
\usepackage[english]{babel}
\usepackage{graphicx}
\usepackage{amsmath, amsfonts, dsfont,amssymb, graphicx, array, tabularx, booktabs,multicol}
\usepackage{amsthm}
\usepackage{epsf,epsfig}
\usepackage{setspace}

\usepackage{mathrsfs}
\usepackage{multirow}
\usepackage{url}
\usepackage{color}
\usepackage[nolist,printonlyused]{acronym}      
\usepackage{comment}
\usepackage{cite}
\usepackage{bm}
\usepackage{tikz}
\usetikzlibrary{decorations.pathreplacing}

\usepackage{hyperref}
\usepackage{mathtools}

\usepackage{blindtext}

\usepackage{stfloats}

\newcommand{\mx}[1]{\mathbf{#1}}

\newcommand{\bs}[1]{\boldsymbol{#1}}

\definecolor{amber}{rgb}{1.0, 0.49, 0.0}
\definecolor{ao}{rgb}{0.0, 0.5, 0.0}

\def\R2#1{\textcolor{black}{#1}}
\def\R3#1{\textcolor{black}{#1}}

\definecolor{copperrose}{rgb}{0.6, 0.4, 0.4}
\definecolor{azure}{rgb}{0.0, 0.5, 1.0}
\definecolor{ashgrey}{rgb}{0.7, 0.75, 0.71}
\definecolor{chestnut}{rgb}{0.8, 0.36, 0.36}
\definecolor{airforceblue}{rgb}{0.36, 0.54, 0.66}
\definecolor{cadmiumorange}{rgb}{0.93, 0.53, 0.18}
\definecolor{bleudefrance}{rgb}{0.19, 0.55, 0.91}
\definecolor{carolinablue}{rgb}{0.6, 0.73, 0.89}
\definecolor{blue(ncs)}{rgb}{0.0, 0.53, 0.74}
\definecolor{dodgerblue}{rgb}{0.12, 0.56, 1.0}
\definecolor{cssgreen}{rgb}{0.0, 0.5, 0.0}
\definecolor{cadmiumgreen}{rgb}{0.0, 0.42, 0.24}
\definecolor{cadmiumorange}{rgb}{0.93, 0.53, 0.18}
\definecolor{amaranth}{rgb}{0.9, 0.17, 0.31}
\definecolor{bluegray}{rgb}{0.4, 0.6, 0.8}
\definecolor{cerulean}{rgb}{0.0, 0.48, 0.65}
\definecolor{ceil}{rgb}{0.57, 0.63, 0.81}
\definecolor{antiquefuchsia}{rgb}{0.57, 0.36, 0.51}
\definecolor{bronze}{rgb}{0.8, 0.5, 0.2}
\definecolor{carrotorange}{rgb}{0.93, 0.57, 0.13}
\definecolor{coolgrey}{rgb}{0.55, 0.57, 0.67}
\definecolor{corn}{rgb}{0.98, 0.93, 0.36}
\definecolor{frenchbeige}{rgb}{0.65, 0.48, 0.36}
\definecolor{dandelion}{rgb}{0.94, 0.88, 0.19}
\definecolor{cadet}{rgb}{0.33, 0.41, 0.47}

\renewcommand{\triangleq}{\mathbin{\setstackgap{S}{0pt}\stackMath\Shortstack{\smalltriangleup\\ =}}}
\usepackage[usestackEOL]{stackengine} 

\title{Convexity Meets Curvature:\\Lifted Near-Field Super-Resolution%
\thanks{This work is supported in part by Digital Futures and the EU Horizon projects MULTIPLY-6G and QUEST-6G under Project IDs 101293106 and 101292676, respectively. G. Fodor was also supported by SSF grant FUS21-0004 SAICOM.}
}

\author{
Sajad Daei$^{1}$, G\'abor Fodor$^{1,2}$, Mikael Skoglund$^{1}$\\
$^{1}$KTH Royal Institute of Technology, Stockholm, Sweden\\
$^{2}$Ericsson Research, Stockholm, Sweden\\
\{sajado,gaborf,skoglund\}@kth.se
}

\begin{document}
\maketitle

\begin{abstract}
Extra-large apertures, high carrier frequencies, and integrated sensing and communications (ISAC) are pushing array processing into the Fresnel region, where spherical wavefronts induce a range-dependent phase across the aperture. This curvature breaks the Fourier/Vandermonde structure behind classical subspace methods, and it is especially limiting with hybrid front-ends that provide only a small number of pilot measurements. Consequently, practical systems need continuous-angle resolution and joint angle–range inference where many near-field approaches still rely on costly 2D gridding. We show that \emph{convexity can meet curvature} via a lifted, gridless super-resolution framework for near-field measurements. The key is a Bessel-Vandermonde factorization of the Fresnel-phase manifold that exposes a hidden Vandermonde structure in angle while isolating the range dependence into a compact coefficient map. Building on this, we introduce a lifting that maps each range bin and continuous angle to a structured rank-one atom, converting the nonlinear near-field model into a linear inverse problem over a row-sparse matrix. Recovery is posed as atomic-norm minimization and an explicit dual characterization via bounded trigonometric polynomials yields certificate-based localization that super-resolves off-grid angles and identifies active range bins. Simulations with strongly undersampled hybrid observations validate reliable joint angle-range recovery for next-generation wireless and ISAC systems.

\end{abstract}


\section{Introduction}
\label{sec:intro}
\textbf{Near-field arrays and loss of Fourier structure:}
The steady growth of array apertures in emerging wireless systems, driven by higher carrier frequencies, tighter spectrum reuse, and the convergence of communication with sensing, is reshaping the operating regime of multi-antenna signal processing. When the aperture becomes electrically large, many scatterers and targets of interest lie in the near field of the array. In this regime, the received field is more accurately described by spherical wavefronts rather than plane waves. As a consequence, the phase variation across the array depends not only on the direction of arrival but also on the propagation distance, and the far-field Fourier model that supports classical beamspace processing no longer applies \cite{lu2014overview,liu2024near,daei2025when} .
This change has practical and conceptual implications. Practically, hybrid analog-digital architectures typically expose only a small number of linear measurements per pilot snapshot, so near-field estimation must succeed in a strongly undersampled regime. Conceptually, near-field responses do not admit a simple Fourier dictionary: the curvature-induced phase profile is range-dependent and interacts with direction in a way that destroys the shift-invariance and Vandermonde properties that many estimators rely on. These challenges are amplified in multi-path environments where several components may be closely spaced in direction yet separated in range (or vice versa), precisely the scenario where super-resolution is most valuable\cite{daei2025timely,daei2025near,tang2013compressed}.

\textbf{Limitations of existing methods:}
A common response to near-field modeling is to discretize both range and angle and apply sparse recovery with a two-dimensional dictionary \cite{cui2022channel,zhang2023near,liu2024near,wang2023near}. While conceptually straightforward, this strategy inherits two well-known difficulties. First, grid mismatch can dominate performance: in high-resolution regimes the true parameters rarely align with the discretization, causing bias and leakage that are difficult to control. Second, the computational burden grows rapidly with grid refinement, and becomes particularly acute when the estimation must be repeated across orthogonal frequency division multiplexing (OFDM) subcarriers or time slots\cite{chi2011sensitivity}.
Beyond 2D gridding, two representative classes of near-field estimators still leave important gaps. 
First, a recent gridless alternative \cite{xi2025near} combines a second-order Taylor approximation with an atomic-norm program; however, its convexification relies on an auxiliary low-dimensional modulation/subspace model that represents the near-field phase variation through an \emph{unknown waveform}. 
This adds a non-physical tuning burden (subspace dimension/structure) not uniquely dictated by the near-field manifold, and it can affect identifiability, estimation accuracy, and computational cost. 
Second, many practical solutions resort to local approximations of the spherical model and iterative nonlinear fitting \cite{xi2025near}. Such methods can be sensitive to initialization, degrade when components are closely spaced, and become fragile under compressed measurements. 
Crucially, neither class provides a unified \emph{certificate-based} mechanism, in the sense of dual-polynomial/convex-geometry optimality conditions-that guarantees stable recovery of continuous parameters under undersampling \cite{tang2013compressed,candes2014towards,chandrasekaran2012convex}.

\textbf{Proposed approach: linearize by lifting, then super-resolve.} 
 This paper develops a different approach that restores the linear inverse-problem structure while respecting near-field physics. The key idea is to convert the near-field array manifold into a representation that is linear in a lifted domain. Instead of forcing the near-field response into a far-field dictionary or discretizing the angular domain, we exploit a harmonic expansion that reveals a hidden Vandermonde structure in angle, coupled to range-dependent coefficients. This factorization enables a lifting procedure that maps each propagation path to a simple structured matrix atom. In the lifted domain, hybrid pilot measurements become linear functionals of a low-complexity superposition of such atoms.
Once the model is linearized, the problem becomes amenable to gridless convex super-resolution. In particular, an atomic norm built from lifted near-field atoms provides a principled convex surrogate for the number of paths, and its dual yields explicit trigonometric-polynomial certificates that localize continuous angles. This connects near-field channel estimation to the broader theory of continuous compressed sensing and super-resolution \cite{tang2013compressed,candes2014towards,chandrasekaran2012convex,daei2025timely,daei2023blind}, while introducing new structure specific to spherical-wave propagation.

\subsection{Contributions}
Our contributions establish a principled bridge between near-field wave physics and gridless convex super-resolution under compressed measurements:
\begin{enumerate}
    \item \textbf{Physics-to-harmonics: a near-field manifold with a hidden Vandermonde core.}
We develop a Bessel-Vandermonde harmonic representation of the Fresnel-region steering vector that isolates range dependence into a compact coefficient map while preserving a Vandermonde structure in angle. This representation explains \emph{how and where} curvature creates additional harmonics beyond the far-field model, and it collapses to the classical far-field steering vector as range goes to infinity.

\item  \textbf{Lifted linearization: from nonlinear Fresnel geometry to a structured linear inverse problem.}
Leveraging the harmonic model, we introduce a lifting that maps each (discretized) range bin and continuous angle to a rank-one, row-sparse matrix atom. This yields a linear measurement model for hybrid-combined pilots, turning near-field joint angle-range estimation into a convex-amenable inverse problem with explicit structure (row sparsity across range and continuous sparsity in angle).
\item \textbf{Gridless recovery with certificates: atomic norms and dual-polynomial localization.}
We pose reconstruction as atomic-norm minimization over the lifted atoms and provide an explicit dual characterization via bounded trigonometric polynomials. The resulting certificate-based peak rule identifies active range bins and \emph{super-resolves} off-grid angles without angular discretization. 

\item \textbf{Amplitude estimation via simple least-squares.}
Given the estimated supports, path gains are obtained in closed-form by solving a simple least-square (LS) optimization which results in a complete complex-amplitude estimate with negligible additional computational cost.

\end{enumerate}

\textit{Notation:}
$\mx I_N$ is the $N\times N$ identity matrix; $\mathbb Z$ and $\mathbb Z_+$ denote the integers and positive integers, respectively; and $\otimes$ denotes the Kronecker product. For $\mx X\in\mathbb C^{M\times N}$, $\operatorname{vec}(\mx X)$ stacks its columns. Matrix and vector entries are denoted by $ A[k,\ell]$ and $ x[k]$, respectively. For paired indices, we fix a bijection $p:\mathcal I\to\{1,\ldots,P\}$, e.g., $\mathcal I=[-I_1,I_1]\times[-I_2,I_2]$, and write $A[n,(\ell,q)]\triangleq A[n,p(\ell,q)]$. The $n$-th entry of the near-field steering vector is denoted by $ a_{\mathrm{NF}}(r,\theta)[n]$. Finally, $J_n(\cdot)$ is the Bessel function of the first kind, $\mx e_n$ is the $n$-th canonical basis vector, and $\langle\cdot,\cdot\rangle$ denotes the Euclidean or Frobenius inner product.

\section{System Model}
\label{sec:system}
\begin{figure}
\captionsetup{font=scriptsize}
    \centering
    \includegraphics[scale=0.27]{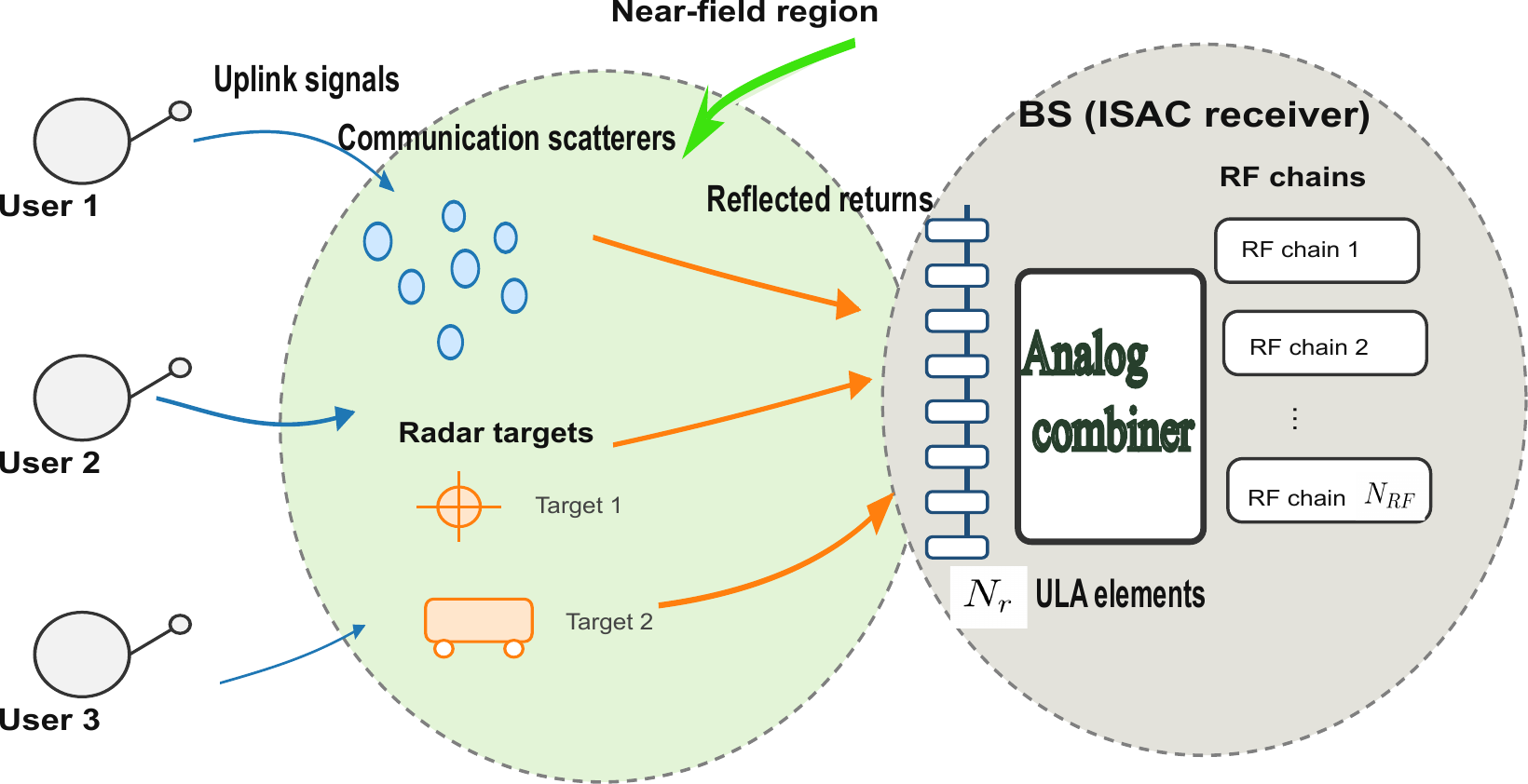}
    \caption{Hybrid XL-MIMO geometry with Analog combiner: a large-aperture BS observes a superposition of spherical-wave paths from multiple single-antenna uplink users and nearby targets/scatterers.}
    \label{fig:systemmodel}
\end{figure}
\begin{figure}
\captionsetup{font=scriptsize}
    \centering
    \includegraphics[scale=0.62]{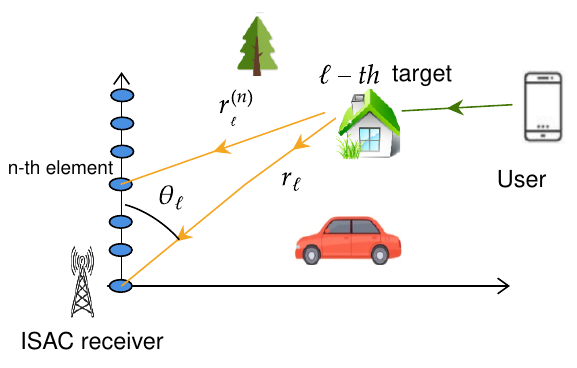}
    \caption{The uplink channel is comprised of $L$ scatterers or targets. Each scatterer has a distinct range and angle parameter relative to the reference antenna with antenna element index $n=0$.}
    \label{fig:isac_mode}
\end{figure}
\subsection{Uplink pilot measurements with hybrid combining}

We consider an uplink time-division duplexing (TDD) extra-large multiple-input multiple-output (XL-MIMO) OFDM system where the base station (BS) employs a uniform linear array (ULA) with $N_r$ antennas with spacing $d=\lambda_c/2$) and $N_{\mathrm{RF}}\ll N_r$ RF chains where $\lambda_c=\tfrac{c}{f_c}$ denotes the carrier wavelength and $f_c$ is the carrier frequency as shown in Figure \ref{fig:systemmodel}. In time slot $p$, the BS applies an analog combiner $\mathbf{B}_p\in\mathbb{C}^{N_{\mathrm{RF}}\times N_r}$ with constant-modulus entries $|{B_p}[i,j]|=\tfrac{1}{\sqrt{N_r}}$\cite{alkhateeb2014channel}. For a given user and subcarrier, let $\mathbf{h}\in\mathbb{C}^{N_r\times 1}$ denote the uplink channel. With pilot symbol $x_p$ (set to $1$ for simplicity), the received measurement is $\mathbf y_p=\mathbf B_p \mathbf h\,x_p+\mathbf B_p\mathbf w_p, \mathbf w_p\sim\mathcal{CN}(\mathbf 0,\sigma^2\mathbf I_{N_r}).$
%
Stacking $P_T$ slots and setting $x_p=1$ gives
\begin{equation}\label{eq:stacked_pilot}
\mathbf y=\mathbf B\mathbf h+\mathbf n\in\mathbb C^{M},
\qquad M\triangleq P_TN_{\mathrm{RF}},
\end{equation}
where
\(
\mathbf B\triangleq[\mathbf B_1^{\top},\ldots,\mathbf B_{P_T}^{\top}]^{\top}
\in\mathbb C^{M\times N_r},
\mathbf n\triangleq
\begin{bmatrix}
\mathbf B_1\mathbf w_1\\[-1mm]
\vdots\\[-1mm]
\mathbf B_{P_T}\mathbf w_{P_T}
\end{bmatrix}.
\)
Thus
\(
\mathbf C_n\triangleq\mathbb E[\mathbf n\mathbf n^{\mathsf H}]
=
\sigma^2\operatorname{blkdiag}
(\mathbf B_1\mathbf B_1^{\mathsf H},\ldots,
\mathbf B_{P_T}\mathbf B_{P_T}^{\mathsf H}).
\)
Let
\(
\mathbf C_n=\sigma^2\mathbf L_{\rm chol}\mathbf L_{\rm chol}^{\mathsf H}.
\)
Pre-whitening yields
\begin{equation}\label{eq:whitened_model}
\mathbf y'=\mathbf B'\mathbf h+\mathbf w',
\end{equation}
with
\(
\mathbf y'=\mathbf L_{\rm chol}^{-1}\mathbf y,\qquad
\mathbf B'=\mathbf L_{\rm chol}^{-1}\mathbf B,\qquad
\mathbf w'=\mathbf L_{\rm chol}^{-1}\mathbf n
\sim\mathcal{CN}(\mathbf 0,\sigma^2\mathbf I_M).
\)

\subsection{Near-field channel model}
\label{sec:channel_model}
We adopt a sparse near-field model with $L$ dominant paths. 
These {paths} may be communications scatterers or radar targets as shown in Figures \ref{fig:systemmodel} and \ref{fig:isac_mode}. The uplink channel can be expressed as:
\begin{align*}
\label{eq:h_nf}
\mathbf h
= \sum_{\ell=1}^{L} c_\ell\,\mathbf a_{\mathrm{NF}}(r_\ell,\theta_\ell)
\qquad c_\ell\in\mathbb C,
\end{align*}
where $(r_\ell,\theta_\ell)$ {denotes} the range and azimuth angle of path $\ell$ w.r.t.\ a reference antenna (antenna $n=0$).
Amplitude variations across antennas are absorbed into $c_\ell$.

Let $d$ be the antenna spacing and index antennas by $n\in\{0,1,\dots,N_r-1\}$.
Define the distance from the scatterer at polar coordinates $(r,\theta)$ to antenna $n$ as
\begin{equation}
\label{eq:rn}
r^{(n)}
\triangleq
\sqrt{r^2+(nd)^2-2r(nd)\cos\theta}.
\end{equation}
We define the near-field steering vector (up to a global phase) as
\begin{align}\label{eq:steer_nf}
\scalebox{.9}{$\mathbf a_{\mathrm{NF}}(r,\theta)[n]
\triangleq
\exp\!\Big(
-jk_{\lambda}\big(r^{(n)}-r\big)
\Big), n=0,\dots,N_r-1,$}
\end{align}
where $k_{\lambda}\triangleq \tfrac{2\pi}{\lambda}$.
The goal is \emph{Given $\mathbf y'$ and $\mathbf B'$, estimate $\{(r_\ell,\theta_\ell,c_\ell)\}_{\ell=1}^{L}$ with continuous angles $\theta_\ell$.}
\section{Bessel-Vandermonde Lifting and Linear Measurement Model}
\label{sec:lift}

We next derive a Bessel-Vandermonde lifting of the near-field steering vector. The lifting converts the nonlinear dependence on range and angle into a linear functional of a higher-dimensional row-sparse matrix.
Define
\(
z_1(n)\triangleq k_\lambda nd,\qquad
z_2(n,r)\triangleq k_\lambda\frac{n^2d^2}{4r}.
\)
For truncation orders $I_1,I_2\in\mathbb Z_+$, let
\(
\mathcal I\triangleq[-I_1,I_1]\times[-I_2,I_2],\quad
P\triangleq|\mathcal I|,\quad
I_{\rm off}\triangleq I_1+2I_2,\quad
N_b\triangleq 2I_{\rm off}+1.
\)
Define
\(
\widetilde{\mx v}(\theta)
\triangleq
\frac{1}{\sqrt{N_b}}
\big[e^{-jI_{\rm off}\theta},\ldots,e^{jI_{\rm off}\theta}\big]^{\mathsf T}
\in\mathbb C^{N_b}.
\)
Let $p:\mathcal I\to\{1,\ldots,P\}$ be a fixed bijection, and define
$\mx S\in\{0,1\}^{P\times N_b}$ by
\(
\mx S[p(\ell,q),\ell+2q+I_{\rm off}+1]=1,
\qquad (\ell,q)\in\mathcal I,
\)
with all other entries equal to zero.
For each $r>0$, define $\mx C(r)\in\mathbb C^{N_r\times P}$ by
\begin{equation}
\label{eq:C_row_thm}
 C(r)[n,(\ell,q)]
\triangleq
e^{-jz_2(n,r)}j^{\ell+q}
J_\ell(z_1(n))J_q(z_2(n,r)).
\end{equation}
Given a range grid $\{r_i\}_{i=1}^{N_d}$, define
\(
\mx D\triangleq
\big[
\operatorname{vec}(\mx C(r_1))\ \cdots\
\operatorname{vec}(\mx C(r_{N_d}))
\big]
\in\mathbb C^{N_rP\times N_d}.
\)
Finally, for $n=0,\ldots,N_r-1$, define
\begin{equation}
\label{eq:Phi_mat}
\bs\Phi_n
\triangleq
\sqrt{N_b}\,
\mx D^{\mathsf T}
(\mx I_P\otimes \mx e_{n+1})
\mx S
\in\mathbb C^{N_d\times N_b}.
\end{equation}

\begin{thm}
\label{thm:lifted}
Consider a ULA with $N_r\ge 2$ antennas, spacing $d$, and antenna indices $n=0,\ldots,N_r-1$. Let
$(r,\theta)\in[r_{\min},r_{\max}]\times(0,\pi)$ with $r_{\min}>0$, and assume that $r$ belongs to the range grid, i.e., $r=r_i$ for some $i$.
Let $\bs\alpha\in\{0,1\}^{N_d}$ be the one-sparse vector satisfying $\bs\alpha[i]=1$, and define
\(
\mx A(r,\theta)
\triangleq
\bs\alpha\,\widetilde{\mx v}(\theta)^{\mathsf H}
\in\mathbb C^{N_d\times N_b}.
\)
Then, for every $n=0,\ldots,N_r-1$,
\begin{equation}
\label{eq:lifted_entry_clean}
 a_{\mathrm{NF}}(r,\theta)[n]
=
\langle \mx A(r,\theta),\bs\Phi_n\rangle
+\delta_n(r,\theta),
\end{equation}
where $\delta_n(r,\theta)
=
\mathcal O\!\left(
k_\lambda\frac{(nd)^3}{r^2}
\right)
+
\delta_n^{\rm tail}(I_1,I_2).$
%
Moreover, 
\begin{align*}
\label{eq:tail_bound}
&|\delta_n^{\rm tail}(I_1,I_2)|
\le
\sum_{|\ell|>I_1}\sum_{q\in\mathbb Z}
|J_\ell(z_1(n))|\,|J_q(z_2(n,r))|+\nonumber\\
&\sum_{|q|>I_2}\sum_{\ell\in\mathbb Z}
|J_\ell(z_1(n))|\,|J_q(z_2(n,r))|.
\end{align*}
\end{thm}
\begin{rem}(Interpretation)
  The theorem says that by \emph{lifting} angle-distance pairs to the rank‑one matrix $\mx{A}$, the inherently nonlinear near‑field response becomes an \emph{affine} functional of $\mx{A}$. This enables convex tools such as atomic‑norm minimization and RIPless guaranties\cite{candes2014towards,tang2013compressed}: recovery reduces to solving a linear inverse problem in the lifted space. $\mx{S}$ is the selection matrix that selects the appropriate harmonics.
We refer to $\mx{A}$ as the ``\textbf{lifted channel matrix}'' and to the linear map $\bm{\varPhi}(\cdot)=\{\bs{\Phi}_n\}_{n=0}^{N_r-1}: \mathbb C^{N_d\times N_b}\to \mathbb C^{N_r\times 1}$  as ``\textbf{Bessel-Vandermonde operator},''  which encodes distance-domain Bessel terms and the angular Vandermonde structure.   
\end{rem}
\begin{rem}[Far-field limit]
\label{rem:ff_standalone}
As $r\to\infty$, $z_2(n,r)\to 0$ so $J_q(z_2(n,r))\to 0$ for $q\neq 0$ and $J_0(z_2(n,r))\to 1$.
Hence only the $q=0$ terms in \eqref{eq:C_row_thm} survive and the approximation reduces to the classical
far-field steering vector
\[
a_{\mathrm{NF}}(r,\theta)[n]\ \to\ a_{\mathrm{FF}}(\theta)[n]
=\exp\!\Big(j k_{\lambda} nd\cos\theta\Big),
\]
recovering the standard Vandermonde/Fourier structure.
\end{rem}
If $I_1\ge \Big\lceil \tfrac{e\pi (N_r-1)d}{\lambda}\Big\rceil$ and $I_2\ge \Big\lceil \tfrac{e}{2}\max_{0\le n\le N_r-1}|z_2(n,r)|\Big\rceil$ (and vanish as $I_1,I_2\to\infty$), then the Bessel truncation tail can be made arbitrarily small. If $r\ge \sqrt{k_\lambda}\,((N_r-1)d)^{\tfrac{3}{2}}$, the the Fresnel remainder can become very small. In practice small truncation orders suffice; see Section \ref{sec:simulation}.

Based on Theorem \ref{thm:lifted} in the Fresnel region and with sufficiently large $I_1, I_2$, we are now able to represent the near-field channel as follows: $h[n]\approx\langle \mx{X},\bs{\Phi}_n\rangle, n=0,..., N_r-1   $
where $\mx{A}_l=\bs{\alpha}_l \widetilde{\mx{v}}^{\mathsf{H}}(\theta_l)$ and $\mx{X}=\sum_{l=1}^L c_l \mx{A}_l$ is an unknown matrix depending on both angle and distance of the scatterers.
The pilot measurements provided in \eqref{eq:whitened_model} can also be expressed in this form:
\begin{align}\label{eq:measure_model}
    &\mx{y}'
    \;=\;
    \mathcal{B}'(\mx{X})
    \;+\;
    \mx{w}'\in\mathbb{C}^{M\times 1},
    \\
    &\bigl(\mathcal{B}'(\mx{X})\bigr)[m]
    \triangleq \langle \mx{X},\bs{\Psi}_m\rangle, m=1,..., M.
\end{align}
where  \(\bs{\Psi}_m \;\triangleq\;\sum_{n=0}^{N_r-1}B'[m,n]\,\bs{\Phi}_n
  \;\in\;\mathbb{C}^{N_d\times N_b}.\)
 The number of unknown variables in the latter lifted model is $N_d N_b$ while the number of known equations is $M\ll N_d N_b$. This makes the system of equations widely under-determined. Since our lifted matrix $\mx{X}$ is actually a superposition of $L\ll \min\{N_d,N_b\}$ rank-one, row-sparse atoms $\alpha_l \widetilde{\mx{v}}^{\mathsf{H}}(\theta_l)$, continuous compressed sensing or super-resolution methods\cite{candes2014towards,tang2013compressed,fernandez2016super} can indeed unlock exact recovery despite $M< N_d N_b$.
To do so, note that
the matrix $\mx{X}$ is composed of few matrix-valued atoms $\mx{A}_l$. Since the distance vector $\bs{\alpha}\in\mathbb{R}^{N_d\times 1}$ is one-sparse, the resultant matrix $\mx{A}=\bs{\alpha}\widetilde{\mx{v}}^H(\theta)$ is both row-sparse and rank one. In fact,
$\mx{A}_l=\bs{\alpha} \widetilde{\mx{v}}^{\mathsf{H}}(\theta)$ is an \emph{atom} or building block from the atomic set 
\begin{align*}
  \scalebox{.9}{$\mathcal{A}=\{\bs{\alpha} \widetilde{\mx{v}}^{\mathsf{H}}(\theta)\in\mathbb{C}^{N_d\times N_b}: \theta\in [0,2\pi), \|\bs{\alpha}\|_1=1,\bs{\alpha}\in\{0,1\}^{N_d\times 1} \}  $}
\end{align*}
Note that since $\bs{\alpha}\in\{0,1\}^{N_d\times 1}$, we have that $\|\bs{\alpha}\|_1=\|\bs{\alpha}\|_0$. To promote row-sparsity and rank-one structure of $\mx{X}$ jointly, we define the following atomic function:
\begin{align*}
 \scalebox{.8}{$\|\mx{X}\|_{\mathcal{A}}=\inf\{t\ge 0: \mx{X}\in t {\rm conv}(\mathcal{A})\}=\inf_{c_l}\{\sum_{l}|c_l|: \mx X=\sum_{l=1}^L c_l \mx A(r_l,\theta_l)\}   $}
\end{align*}
where $\|\mx{X}\|_{\mathcal{A}}$ is called the \emph{atomic} norm \cite{chandrasekaran2012convex,candes2014towards,tang2013compressed} and is the closest convex surrogate to ``number of lifted atoms'', encouraging few active range bins and a sparse continuous angular spectrum per bin.
By minimizing this function subject to satisfying some measurement constraints, one can recover this matrix. This is done in the following optimization:
\begin{align}\label{prob:primal_prob}
    &\min_{\mx{Z}\in\mathbb{C}^{N_d\times N_b}} \|\mx{Z}\|_{\mathcal{A}} ~~ {\rm s.t.}\|\mx y'-\mathcal{B}'(\mx{Z})\|_2\le \eta    
\end{align}
where $\eta$ is an upper-bound for $\|\mx{w}'\|_2$ in \eqref{eq:whitened_model}.
The dual of the latter optimization  problem can be obtained as follows (see e.g. \cite{daei2025timely,daei2023blind,daei2023blind_wiopt,safari2021off,maskan2023demixing,seidi2022novel}):
\begin{align}\label{prob.dual}
&\max_{\mx{q}} {\rm Re} \langle \mx{y}',\mx{q} \rangle  -\eta \|\mx{q}\|_2~{\rm s.t.}~ \|\mathcal{B}'^{\rm Adj}(\mx{q})\|_{\mathcal{A}}^{\text{d}}\le 1,
\end{align}
where $\|\mx X\|^{\rm d}_{\mathcal{A}}\triangleq \sup_{\|\mx Z\|_A\le 1} \langle \mx X,\mx Z\rangle=\sup_{\substack{\theta\in(0,\pi)\\ \|\bs \alpha\|_1=1}} \langle \bs \alpha \widetilde{\mx{v}}^{H}(\theta), \mx Z\rangle=\sup_{\theta}\|\mx Z^H \widetilde{\mx v}(\theta)\|_{\infty}$ is called the dual atomic norm and 
$
\mathcal B'^{\!\rm Adj}(\mx q) =\ \sum_{m=1}^M q_m\,\bs \Psi_m\ \in\ \mathbb{C}^{N_d\times N_b},
$ is the adjoint operator of $\mathcal{B}'(\cdot)$. The following lemma provides a way to state the dual problem \eqref{prob.dual} in an equivalent tractable semidefinite programming (SDP) form. 

\begin{lem}\label{lem.sdp}
Let $\bs{\Psi}_m\triangleq \begin{bmatrix}
{\bs{\psi}_m^1},..., {\bs{\psi}_m^{N_d}}    
\end{bmatrix}^{\top}\in\mathbb{C}^{N_d\times N_b}$ and $\widetilde{\bs{\Psi}}_i=\begin{bmatrix}
{\bs{\psi}}_1^1&\hdots&{\bs{\psi}}_M^i    
\end{bmatrix}\in\mathbb{C}^{N_b\times M} \forall i=1,..., N_d$.  Then, the dual optimization problem \eqref{prob.dual} can be relaxed to the following SDP optimization problem 
\begin{align}\label{prob.sdp}
&\max_{\mx{Q}_i\in\mathbb{C}^{N_b\times N_b},\mx{q}\in\mathbb{C}^{M\times 1}} {\rm Re} \langle \mx{y}',\mx{q}\rangle-\eta \|\mx{q}\|_2\nonumber\\
&{\rm s.t.}~\begin{bmatrix}
\mx{Q}_i&  \widetilde{\bs{\Psi}}_i\mx{q}\\
\mx{q}^H\widetilde{\bs{\Psi}}_i^{\mathsf{H}}&1
\end{bmatrix}\succeq \mx{0}\forall i=1,..., N_d,\\
&\langle \bs{\Theta}_{n},\mx{Q}_i\rangle=1_{n=0}, n=-N_b+1,..., N_b-1,
\end{align}
where $\bs{\Theta}_n$ is the elementary Toeplitz matrix with ones along its $n$-th diagonal and zero elsewhere. 
\end{lem}
Define range-indexed dual polynomials $p_i(\theta)
\triangleq
e_i^{\mathsf{T}}\mathcal B'^{\mathrm{Adj}}(\mx q)\,\widetilde{\mathbf v}(\theta),
\qquad i=1,\dots,N_d.$
At optimum, peaks of $|{p_i(\theta)}|$ identify the active range bins and continuous angles as shown in Figure \ref{fig:dualpol1}. 
Moreover, the support is recovered by
\(
\widehat{\mathcal S}
=
\{(i,\theta):|{p_i(\theta)}|=1\}
=
\{(i_\ell,\theta_\ell)\}_{\ell=1}^{\hat{L}}.
\)
Insert the estimated support $\widehat{\mathcal S}={(\hat r_\ell,\hat\theta_\ell)}_{\ell=1}^{\hat L}$ into \eqref{eq:measure_model} and estimate the gains by an LS debiasing step (cf. \cite[Eq.~29]{daei2025timely}). Define $\mx G\in\mathbb C^{M\times \hat L}$ with
\(
G[m,\ell]\triangleq \langle \mx A(\hat r_\ell,\hat\theta_\ell),\bs\Psi_m\rangle_F,
\)
then
\(
\hat{\bs c}=\arg\min_{\bs c}\|\mx y'-\mx G\bs c\|_2^2
= (\mx G^{\mathsf H}\mx G)^{-1}\mx G^{\mathsf H}\mx y'.
\)



\section{Numerical Experiments}\label{sec:simulation}
\begin{figure}[t]
\captionsetup{font=scriptsize}
    \centering
    \begin{subfigure}[t]{0.48\linewidth}
        \centering
        \includegraphics[scale=.25]{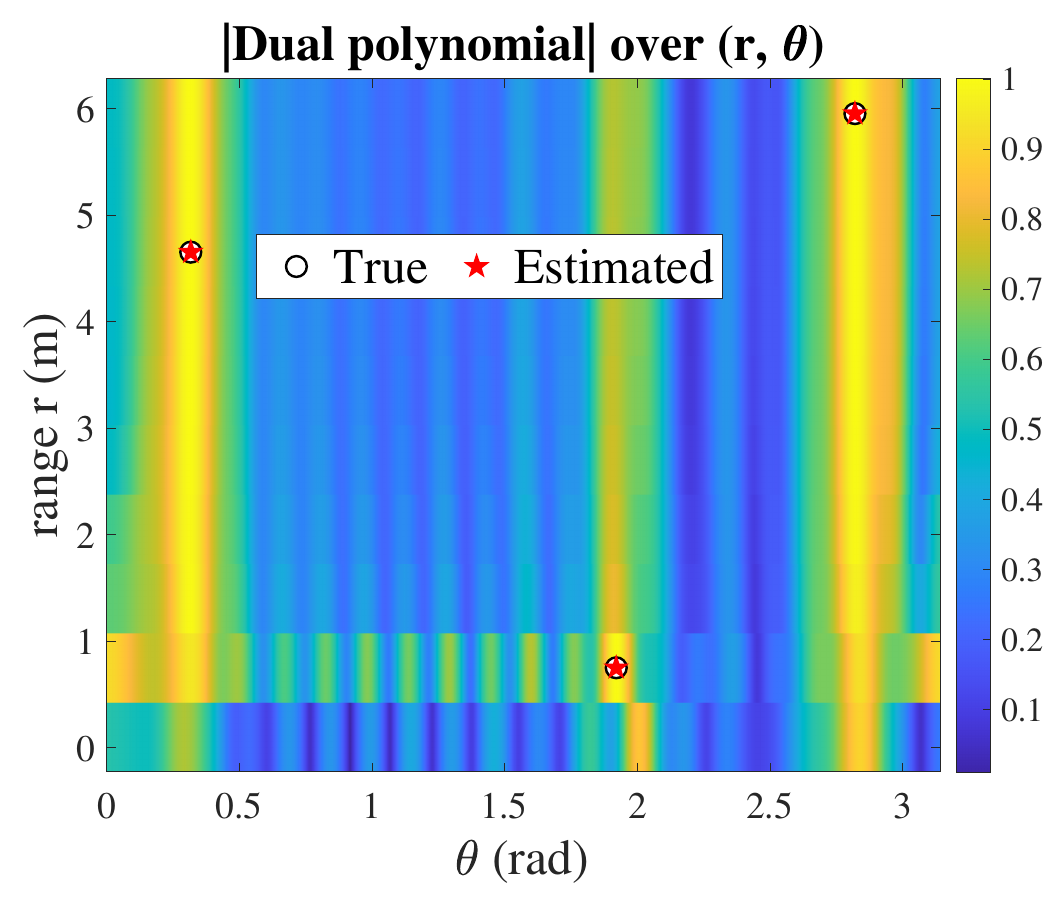}
        \caption{}
        \label{fig:dualpol1}
    \end{subfigure}\hfill
    \begin{subfigure}[t]{0.48\linewidth}
        \centering
        \includegraphics[scale=.25]{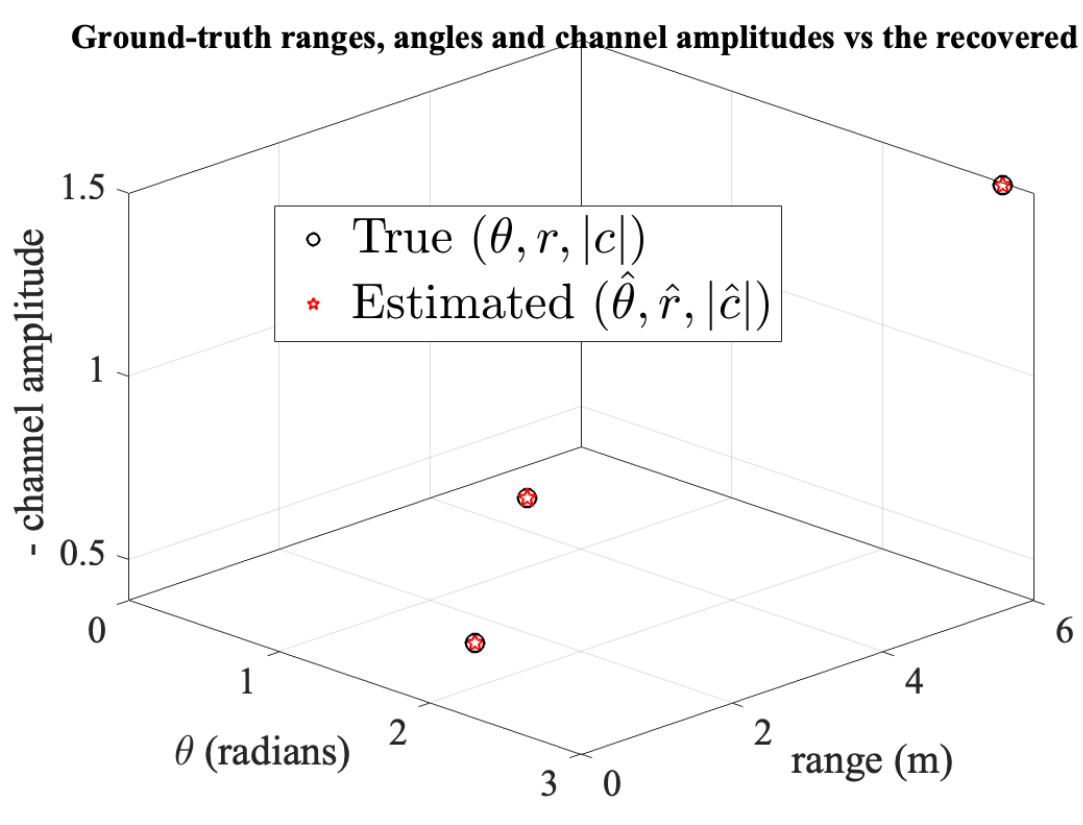}
        \caption{}
        \label{fig:amplitude1}
    \end{subfigure}
    \caption{Gridless recovery outputs: (a) dual polynomial used for localization, (b) recovered amplitudes. The estimates $\widehat{r}_l,\widehat{\theta}_l$ and $\widehat{c}_l$ are precisely aligned with the ground-truth ones.}
    \label{fig:dualpol_amp}
\end{figure}

We illustrate (i) the truncated Bessel-Vandermonde lifting and (ii) the gridless dual-certificate localization.
A uniform linear array with $N_r=64$ antennas and spacing $d=\tfrac{\lambda}{2}$ is simulated at $f_c=100$\,GHz. We use a range grid of
$N_d=10$ bins uniformly spanning $[r_{\min},r_{\max}]=[0.1,6]$\,m and generate $L=3$ near-field paths with
ranges selected on the grid, angles $\theta_\ell\sim{\rm Unif}(0,\pi)$, and different complex gains.
The recovery experiment uses the truncated Bessel-Vandermonde model with $(I_1,I_2)=(20,1)$. The truncation orders control the approximation accuracy, and larger values of $I_1$ and $I_2$ reduce the corresponding Bessel tail terms, as predicted by Theorem \ref{thm:lifted}.
To demonstrate gridless recovery, we solve the dual SDP in Lemma~\ref{lem.sdp} using CVX/SDPT3 and evaluate the range-indexed dual polynomials on a fine angular grid. Peaks of $|p_i(\theta)|$ identify the active range bins and continuous angles, as shown in Fig.~\ref{fig:dualpol_amp}(a). Given the estimated supports, the complex path gains are obtained by a least-squares fit using the reconstructed atoms, as shown in Fig.~\ref{fig:dualpol_amp}(b). The recovered range, angle, and path-gain estimates are consistent with the ground truth for the reported realization.


\section{Conclusion}
This work delivers a simple message for near-field XL-MIMO/ISAC: losing Fourier structure does not require 2D gridding or fragile nonlinear fitting. By exposing a \emph{hidden} Vandermonde core in the Fresnel manifold via a Bessel-Vandermonde expansion and lifting range-angle pairs to structured rank-one atoms, we turn near-field geometry into a linear inverse problem that is compatible with compressed pilots. This is more than a modeling convenience: it enables a fully gridless convex program whose dual polynomials act as {verifiable certificates} for localization. Continuous angles are super-resolved while active range bins are revealed by row support; path gains follow from closed-form least-squares post-processing.
Beyond robust performance under strong undersampling, the framework provides a unifying view across propagation regimes: as curvature vanishes, it reduces to classical far-field super-resolution. The resulting Bessel-Vandermonde operator and certificate-based localization rule form a reusable building block for near-field signal processing, including wideband and multi-user extensions and joint communication-sensing scenarios, where hybrid hardware makes each measurement highly valuable.


\section{Proof of Theorem \ref{thm:lifted} }\label{proof.thm.lifted}

Let \(r^{(n)}=r\sqrt{1+u},
u\triangleq \Big(\tfrac{nd}{r}\Big)^2-2\Big(\tfrac{nd}{r}\Big)\cos\theta.\)
For fixed $r\ge r_{\min}$ and $n\le N_r-1$, we have $|u|\le 2\tfrac{nd}{r}+\big(\tfrac{nd}{r}\big)^2$.
Using Taylor's theorem for $\sqrt{1+u}$ around $u=0$, i.e., $\sqrt{1+u}=1+\tfrac{u}{2}-\tfrac{u^2}{8}+\mathcal{O}(|u|^3)
$, we have
\begin{align*}
\scalebox{.9}{$r^{(n)}-r
=r\Big(\tfrac{u}{2}-\tfrac{u^2}{8}\Big)+r\,\rho(u)
=
-nd\cos\theta+\tfrac{n^2d^2}{2r}\sin^2\theta+\varepsilon_n(r,\theta),$}
\end{align*}
with $|\varepsilon_n(r,\theta)|=\mathcal{O}(\tfrac{(nd)^3}{r^2})$.
Using $\sin^2\theta=\tfrac{1-\cos(2\theta)}{2}$ yields the standard Fresnel form \(r^{(n)}-r=-nd\cos\theta+\tfrac{n^2d^2}{4r}(1-\cos(2\theta))+\varepsilon_n(r,\theta).
\)
Substituting into the definition of $a_{\mathrm{NF}}$ gives
\begin{align}
&\scalebox{.7}{$a_{\mathrm{NF}}(r,\theta)[n]=\exp\!\Big(j z_1(n)\cos\theta\Big)\,
\exp\!\Big(-j z_2(n,r)\big(1-\cos(2\theta)\big)\Big)\,
\exp\!\big(-jk_\lambda\varepsilon_n(r,\theta)\big).$}
\label{eq:NF_fact}
\end{align}
Define the {Fresnel-approximated} entry by dropping the last factor:
\[
a_{\mathrm{F}}(r,\theta)[n]\triangleq
\exp\!\Big(j z_1(n)\cos\theta\Big)\,
\exp\!\Big(-j z_2(n,r)\big(1-\cos(2\theta)\big)\Big).
\]
Then, using $|e^{-jt}-1|\le |t|$, $|a_{\mathrm{NF}}-a_{\mathrm{F}}|
=
|a_{\mathrm{F}}|\cdot|e^{-jk_\lambda\varepsilon_n}-1|
\le k_\lambda|\varepsilon_n(r,\theta)|$
which proves the Fresnel remainder is $\mathcal{O}(k_{\lambda}\tfrac{(nd)^3}{r^3})$.
Rewrite the quadratic term as
\[
\exp\!\Big(-j z_2(1-\cos(2\theta))\Big)=e^{-jz_2}\exp\!\big(jz_2\cos(2\theta)\big),
\]
where $z_2\triangleq z_2(n,r)$.
By the Jacobi-Anger identity,
\begin{align*}
&e^{j z_1\cos\theta}=\sum_{\ell\in\mathbb Z} j^\ell J_\ell(z_1)e^{j\ell\theta}, e^{j z_2\cos(2\theta)}=\sum_{q\in\mathbb Z} j^q J_q(z_2)e^{j2q\theta}.
\end{align*}
Multiplying and using \eqref{eq:NF_fact} (without the Fresnel remainder factor) yields the exact series
\begin{align}
\scalebox{.85}{$a_{\mathrm{F}}(r,\theta)[n]
=
e^{-jz_2(n,r)}\sum_{\ell\in\mathbb Z}\sum_{q\in\mathbb Z}
j^{\ell+q}J_\ell(z_1(n))J_q(z_2(n,r))e^{j(\ell+2q)\theta}.$}
\end{align}
Let $s_n(r,\theta)$ denote the truncated sum over $|\ell|\le I_1$, $|q|\le I_2$, and define the tail
$\delta_n^{\mathrm{tail}}\triangleq a_{\mathrm{F}}-s_n$.
Define $\mx v(\theta)\in\mathbb C^{P\times 1}$ with entries $v(\theta)[(\ell,q),1]=e^{j(\ell+2q)\theta}$ for
$|\ell|\le I_1$, $|q|\le I_2$. By construction of $\mx S$ and $\widetilde{\mx{v}}(\theta)$,
\begin{align}
\label{eq:v_S_vtilde}
\mx v(\theta)=\sqrt{N_b}\,\mx S\,\widetilde {\mx v}(\theta).
\end{align}
Also, by the definition of $\mx C(r)$ in \eqref{eq:C_row_thm}, the truncated sum can be written as
\begin{align}
   \scalebox{.9}{$ s_n(r,\theta)=\sum_{|\ell|\le I_1}\sum_{|q|\le I_2} C(r)[n,(\ell,q)]\,v(\theta)[(\ell,q),1]
= \mx c_n(r)^{\mathsf T}\mx v(\theta),$}
\end{align}
where $\mx c_n(r)^{\mathsf T}$ is the $n$-th row of $\mx C(r)$.
Using the standard row-extraction identity for column-wise vectorization,
\[
\mx c_n(r)=\big(\mx I_P\otimes \mx e_{n+1}^{\mathsf T}\big)\,\mathrm{vec}(\mx C(r)).
\]
On the range grid, $\mathrm{vec}(\mx C(r))=\mx D\bs \alpha$ by definition of $\mx D$ and the one-sparsity of $\bs \alpha$.
Combining with \eqref{eq:v_S_vtilde} gives
\begin{align*}
s_n(r,\theta)
&= \mathrm{vec}(\mx C(r))^{\mathsf T}\big(\mx I_P\otimes \mx e_{n+1}\big)\,\sqrt{N_b}\,\mx S\,\widetilde{\mx{v}}(\theta) \\
&= \bs \alpha^{\mathsf T}\mx D^{\mathsf T}\big(\mx I_P\otimes \mx e_{n+1}\big)\mx S\,\sqrt{N_b}\,\widetilde{\mx{v}}(\theta)
= \bs \alpha^{\mathsf T}\bs \Phi_n\,\widetilde{\mx{v}}(\theta),
\end{align*}
where $\bs \Phi_n$ is as in \eqref{eq:Phi_mat}.
Finally, for $\mx A(r,\theta)=\bs \alpha\,\widetilde{\mx{v}}(\theta)^{\mathsf H}$,
\begin{align*}
\scalebox{.8}{$\langle \mx A(r,\theta),\bs \Phi_n\rangle
=\mathrm{Tr}\!\big(\mx A(r,\theta)^{\mathsf H}\bs{\Phi}_n\big)
=\mathrm{Tr}\!\big(\widetilde{\mx v}(\theta)\bs{\alpha}^{\mathsf T}\bs \Phi_n\big)
=\bs{\alpha}^{\mathsf T}\bs \Phi_n\,\widetilde{\mx v}(\theta)
=s_n(r,\theta).$}
\end{align*}
Thus $ a_{\mathrm{NF}}[n]=s_n+\delta_n^{\mathrm{F}}+\delta_n^{\mathrm{tail}}$, which is exactly \eqref{eq:lifted_entry_clean}.

\bibliographystyle{IEEEtran}
\bibliography{refs}

\end{document}